\documentstyle[12pt,epsf]{article}

\begin{document}

\begin{titlepage}

\begin{flushright}
\vspace{-2.0cm}{\normalsize UTHEP-300\\
April 1995\\}
\end{flushright}

\vspace*{2.0cm}

\begin{centering}
{\Large \bf Chiral Phase Transition 
in Lattice QCD with Wilson Quarks
}

\vspace{1cm}

{\large Y.\ Iwasaki,$^{\rm a}$ K.\ Kanaya,$^{\rm a}$
S.\ Sakai,$^{\rm b}$ and T.\ Yoshi\'e$^{\rm a}$
}

\vspace{0.2cm}

{\it
$^{\rm a}$ 
Institute of Physics, University of Tsukuba, 
Ibaraki 305, Japan\\
$^{\rm b}$ 
Faculty of Education, Yamagata University, 
Yamagata 990, Japan
}

\end{centering}

\vspace{1.5cm}\noindent
The nature of the chiral phase transition in lattice QCD 
is studied for the cases of 2, 3 and 6 flavors with 
degenerate Wilson quarks, mainly on a lattice with the 
temporal direction extension $N_t=4$.  We find that the 
chiral phase transition is continuous for the case of 2 
flavors, while it is of first order for 3 and 6 flavors.

\vfill \noindent

\end{titlepage}


\section{Introduction}
In this article we investigate the nature
of the chiral phase transition in lattice QCD
for various numbers of flavors $N_F$ in the case of 
degenerate quarks,
taking the Wilson formalism of fermions on the lattice.
The determination of the order of the chiral transition, 
in particular for $N_F=2$, is a first
step toward the understanding of what happens in the QCD 
phase transition,
which is supposed to occur at the early Universe and 
possibly at heavy ion collisions.
It is also important to compare
the results for various number of flavors
with theoretical predictions.
Because the Wilson formalism of fermions on the lattice
is the only known formalism which possesses a local action 
for any number of flavors, it is urgent to investigate the
finite temperature chiral phase transition for various 
number of flavors with Wilson quarks.

For each $N_F$ there are two parameters:
$\beta$ ($=6/g^2$; g is the gauge coupling constant) and
the hopping parameter $K$.
Because chiral symmetry is explicitly broken in the Wilson
formalism even for a vanishing bare quark mass,
we have to begin with the identification of the chiral limit.
We define the quark mass through the axial Ward 
identity\cite{Bo,ItohNP}
$$
2 m_q <\,0\,|\,P\,|\,\pi\,>= - m_\pi <\,0\,|\,A_4\,|\,\pi\,>
$$
where $P$ is the pseudoscalar density and $A_4$ the fourth 
component of the local axial vector current.
At zero temperature
the pion mass vanishes at the point which is
almost identical to that where the quark mass 
vanishes\cite{ItohNP,mm}.
It has been further shown that
the value of the quark mass does not depend on
whether the system is in
the high or the low temperature phase through simulations
at $\beta=5.85$ in the quenched QCD \cite{ChQ} and
at $\beta=5.5$ for the $N_F=2$ case\cite{ChNf2}.
Therefore we can identify the chiral limit by the 
vanishing point of the quark mass $m_q$.
This defines a curve $K_c(\beta)$ in the $\beta$ -- $K$ plane.
Alternatively we may use the vanishing point of $m_\pi^2$ 
in the confining phase.

At small $\beta$ region ($\beta \le 5.0$)
where we have made simulations in this work,
$m_q$ in the deconfining phase
does not agree with that in the confining phase
and the proportionality between the $m_q$ in the deconfining 
phase and $m_\pi^2$ in the confining phase is lost,
contrary to the situation for $\beta \ge 5.5$ mentioned above.
We interpret this as a lattice artifact.
On the other hand, the proportionality between the $m_q$ in the 
confining and $m_\pi^2$ in the confining phase is well satisfied.
Therefore we adopt as $K_c(\beta)$ the vanishing point of $m_q$ 
in the confining phase $K_c(m_q)$ or that of $m_\pi^2$ in the 
confining phase $K_c(m_\pi^2)$.
The values of $K_c(\beta)$ show a slight dependence 
on the choice of $K_c(m_q)$ or $K_c(m_\pi^2)$.
They also slightly depend on $N_F$ and $N_t$.
Those at $\beta=4.5$ are listed in Table \ref{tab:1}.
We find that the difference due to $N_F$ is of the same order of
magnitude as that due to the difference of definition.
The $K_c(m_\pi^2)$ for $N_F=2$ for various $\beta$'s are 
listed in Table \ref{tab:2}.

The temperature on a lattice is given by $1/N_t a$, where
$N_t$ is the lattice size in the temporal direction and
$a$ is the lattice spacing. 
The location of the transition or crossover $K_t(\beta)$ from 
the high temperature regime to the low
temperature regime, for a given $\beta$
and a fixed temporal lattice size $N_t$,
is identified
by a sudden change of
physical observables such as the plaquette, the Polyakov 
line and screening hadron spectrum.

The chiral transition occurs at the crossing point of the 
$K_t(\beta)$ and $K_c(\beta)$ line.
However, whether the transition $K_t$ line with a fixed $N_t$
crosses the $K_c$ line is not trivial.
As was first noticed by  Fukugita {\it et al.}\cite{FOU},
the  $K_t$ line creeps deep into the strong coupling region.
Therefore before discussing the order of the transition, 
we have to address ourselves to the problem
whether the chiral limit of the finite temperature transition 
exists at all.

In a previous paper\cite{strong} 
we showed that when $N_F \ge 7$, the $K_t$ line
does not cross the $K_c$ line at finite $\beta$. 
In this article we identify
the crossing point for the case $N_F \le 6$ and determine 
the order of the transition. Preliminary reports are given 
in \cite{preliminary}. 
We take the strategy of
performing simulations on the line $K_c(\beta)$ starting from
a $\beta$ in the high temperature phase
and reducing $\beta$.
We call this method ``on-$K_c$'' simulations.
The number of iteration $N_{\rm inv}$ needed for the quark 
matrix inversion, in general, provides a good indicator
to discriminate the high temperature phase from the low 
temperature phase. 
The use of $N_{\rm inv}$ as an indicator
is extremely useful ``on-$K_c$'',
because $N_{\rm inv}$ is enormously large on the $K_c$ line 
in the confining phase,
while it is of order several hundreds in the deconfining phase.
This difference is due to the fact that
there are zero modes around $K_c$ in the low temperature phase,
while none exists in the high temperature 
phase\cite{strong,zero_mode}.
Therefore there is a sudden drastic change of $N_{\rm inv}$ 
across the boundary of the two phases.
Combining this with measurements
of the Polyakov loop, the plaquette and hadron screening masses,
we identify the crossing
point $\beta_{ct}$ of the $K_t(\beta)$ line with the $K_c(\beta)$ 
line.
We also check that the crossing point thus determined is 
consistent with an extrapolation of the line $K_t(\beta)$ 
toward the chiral limit.
From the behavior of physical quantities toward $\beta_{ct}$,
we are able to determine the order of the chiral transition.

\section{Simulations}
We mainly perform simulations for the case of $N_t=4$ 
with the spatial size $8^2 \times 10$.
To study the $N_t$ dependence for the $N_F=2$ case, 
we also make simulations at $N_t=6$
and at $N_t=18$ with $12^3$ and $18^2 \times 24$ spatial 
lattices, respectively.
We generate gauge configurations 
by Hybrid Monte Calro (HMC) algorithm for $N_F=2$ with
a molecular dynamics step chosen in such a way that the 
acceptance rate is
about 80 -- 90\%, while for $N_F \ge 3$
by the hybrid R algorithm with the molecular
dynamics step $\Delta \tau =0.01$, unless otherwise stated.
The inversion of the quark matrix 
is made by a minimal residual method or a conjugate gradient 
(CG) method.
When the hadron spectrum is calculated,
the lattice is duplicated in the direction of lattice size 
10 or 12.
We use an anti-periodic boundary  condition for quarks in the 
$t$ direction and periodic boundary conditions otherwise.
The statistics is in general total $\tau=$ several hundreds, 
and the plaquette and the Polyakov loop are measured every 
simulation time unit and
hadron spectrum is calculated every $\delta\tau=10$ (or less 
depending on the total statistics). When the value of $\beta$ 
is small, the fluctuation of
physical quantities are small\cite{strong} 
and therefore we think the statistics
is sufficient for the purpose of this article.

\section{Results}
\subsection{$N_F=2$}
\begin{figure}
\centerline{ \epsfxsize=13cm \epsfbox{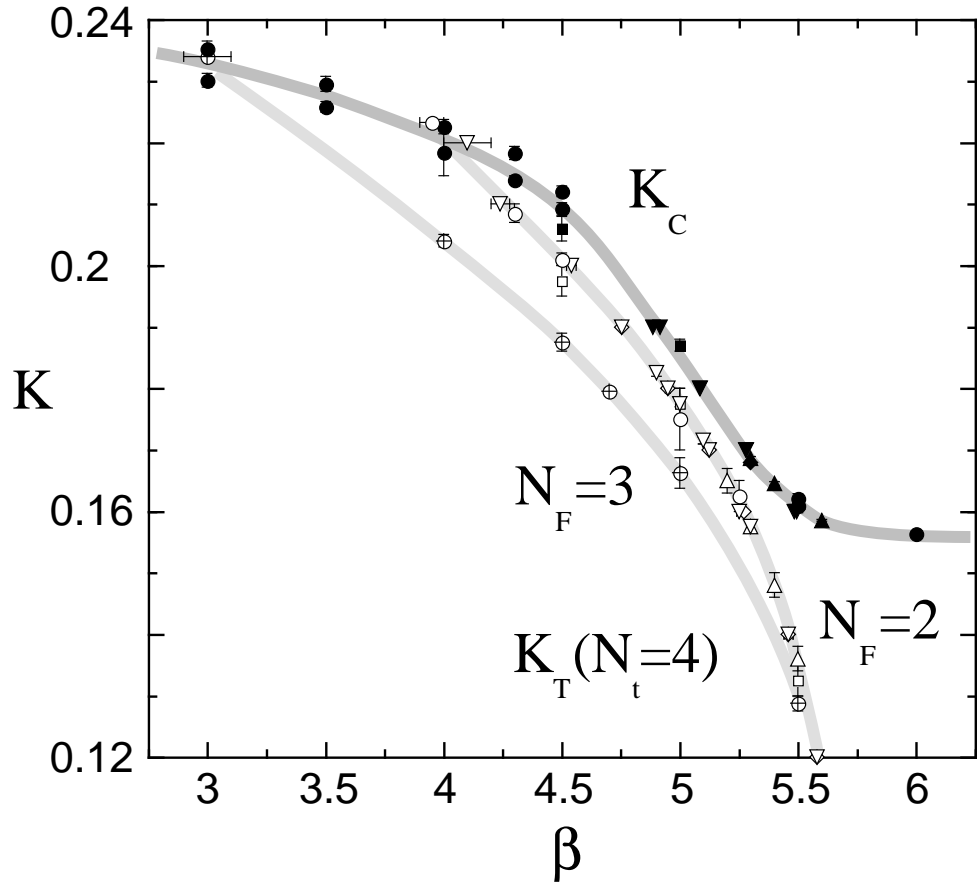} }
\caption{{
Phase diagram for $N_F=2$ and 3.
Filled symbols are for $K_c(m_{\pi}^2)$ and $K_c(m_q)$.
Open symbols are for $K_t(N_t=4)$ for $N_F=2$ while
open circles with cross for $N_F=3$. 
Circles are our data. 
Lines are for guiding eyes. 
}
\label{figF2F3KcKt}}
\end{figure}

\begin{figure}
\centerline{ \epsfxsize=13cm \epsfbox{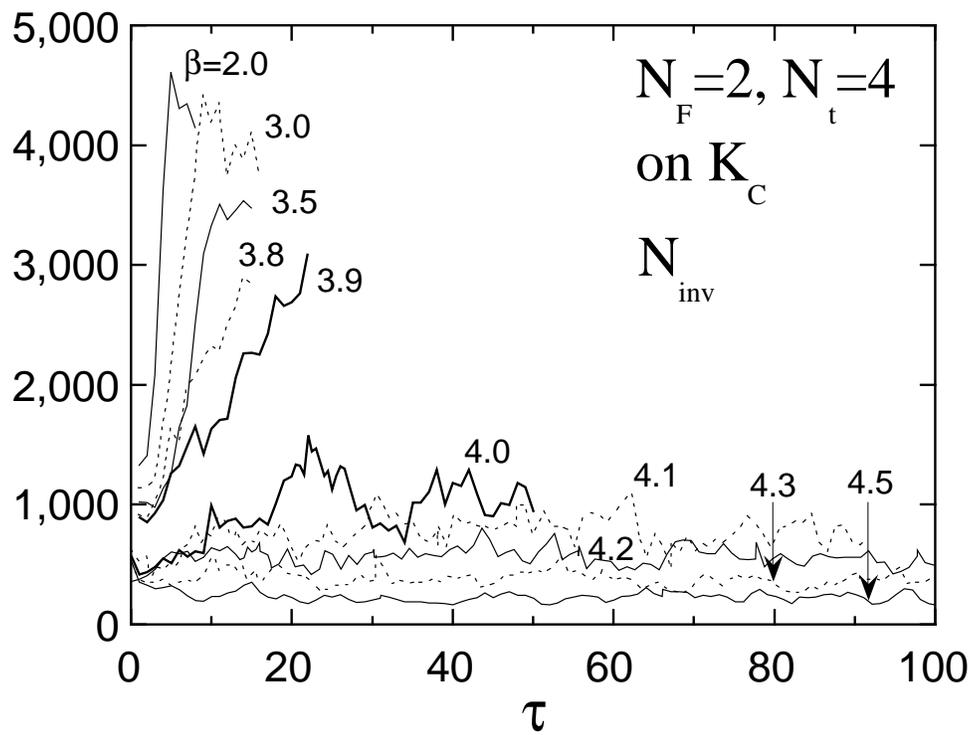}}
\caption{{
Molecular-dynamics time history of $N_{\rm inv}$ for 
$N_F=2$ on $K_c$. 
}
\label{figH2Ninv}}
\end{figure}

Now let us discuss the results
for the $N_F=2$  case at $N_t = 4$. 
The $K_t(\beta)$'s obtained by various 
groups\cite{various,preliminary}
are plotted in Fig.~\ref{figF2F3KcKt} 
together with the $K_c(\beta)$ line\cite{kc}.
In order to confirm the existence of the crossing point
we take the largest (farthest)
$K_c$, that is $K_c(m_\pi^2)$ for $N_F=2$ in Table \ref{tab:2}
and interpolated ones for ``on-$K_c$'' simulations.
We first perform ``on-$K_c$'' simulations 
by the R algorithm to identify
the crossing point, because 
it is very time consuming due to low acceptance rate
for the HMC algorithm
on $K_c$ in the confining phase.
We find that when $\beta \ge 4.0$, $N_{\rm inv}$ stays around 
several hundreds,
while $\beta \le 3.9$ it increases with $\tau$ and exceeds 
several thousands (see Fig.~\ref{figH2Ninv}).
Therefore we identify the crossing point
at $\beta_{ct}\sim 3.9$ -- 4.0.
This $\beta_{ct}$ is
consistent with a linear extrapolation of the $K_t$ line as
is shown in Fig.~\ref{figF2F3KcKt}.
Our results for $K_t$ is summarized in Table \ref{tab:3}.

\begin{figure}
\centerline{ \epsfxsize=13cm \epsfbox{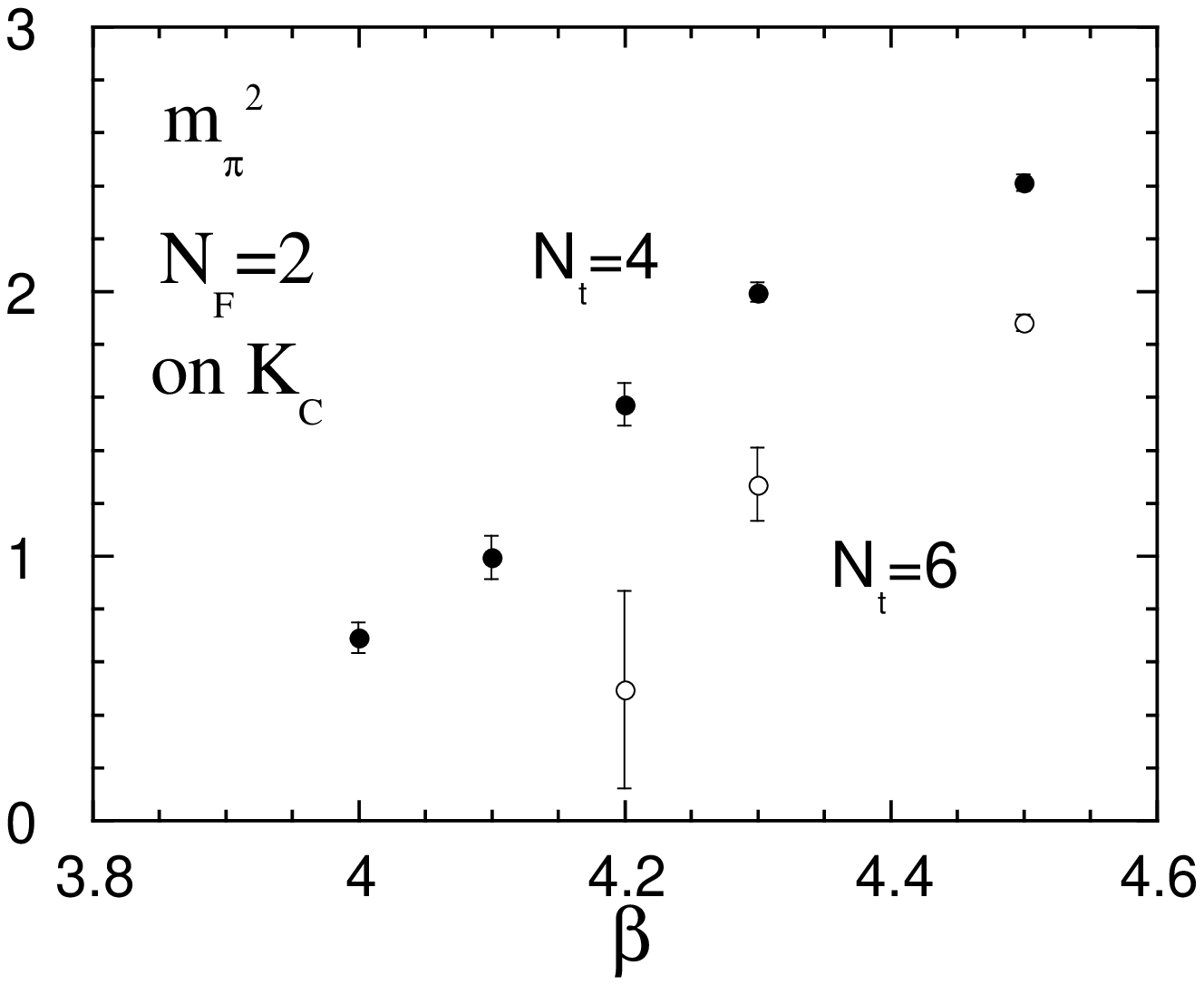}}
\caption{{
$m_{\pi}^2$ for $N_F=2$ on $K_c$. 
}
\label{figF2Mpi}}
\end{figure}

Then we repeat ``on-$K_c$'' simulations by the HMC algorithm
for $\beta \ge 4.0$
in order to measure physical observables. $N_{\rm inv}$ for 
$\beta \ge 4.0$ plotted in Fig.~\ref{figH2Ninv} is for the HMC 
algorithm, which is similar to that for
the R algorithm.
The $\Delta\tau$ toward $\beta_{ct}$ should be taken small 
in order to keep the acceptance rate reasonably high 
(for $\beta=4.0$, 4.1 and 4.2 we use $\Delta\tau=0.002$, 0.005
and 0.005 to get acceptance rates 0.91, 0.79 and 0.93, 
respectively).
The value of $m_{\pi}^2$ thus obtained
decreases toward zero as the chiral transition is approached 
(see Fig.~\ref{figF2Mpi}).

We find no two-state signals around $\beta_{ct}$.
This is in sharp contrast with the $N_F=3$ and 6 cases
where we find clear two-state signals at $\beta_{ct}$,
as discussed below.
This together with the vanishing $m_{\pi}^2$ toward $\beta_{ct}$
indicate that the chiral phase transition is continuous for 
$N_F=2$.

The results by ``on-$K_c$'' simulations on the $N_t=6$ lattice
are similar to those on the $N_t=4$ lattice. 
The estimated transition point is $\beta_{ct}\sim $ 4.0 -- 4.2.
The value of $m_{\pi}^2$ plotted in Fig.~\ref{figF2Mpi}
again decreases toward zero as $\beta$ approaches $\beta_{ct}$.
For $N_t=18$ with the spatial size $18^2 \times 24$,
we find that the transition is at
$\beta_{ct} \sim 4.5$ -- 5.0.
Although the spatial size is not large enough, this result 
suggests that the shift of $\beta_{ct}$ with $N_t$ is very slow.

\subsection{$N_F=3$}
Now we perform ``on-$K_c$'' simulations for $N_F=3$.
In order to confirm the existence of the crossing point
we take the largest (farthest)
$K_c$, that is $K_c(m_\pi^2)$ for $N_F=2$
at $\beta$'s we have studied,
since this is the most stringent condition for the existence 
of $\beta_{ct}$.
We use them and interpolated ones
here, and also for $N_F=6$ interpolated ones with $K_c=0.25$ 
at $\beta=0$.

\begin{figure}
\centerline{ \epsfxsize=13cm \epsfbox{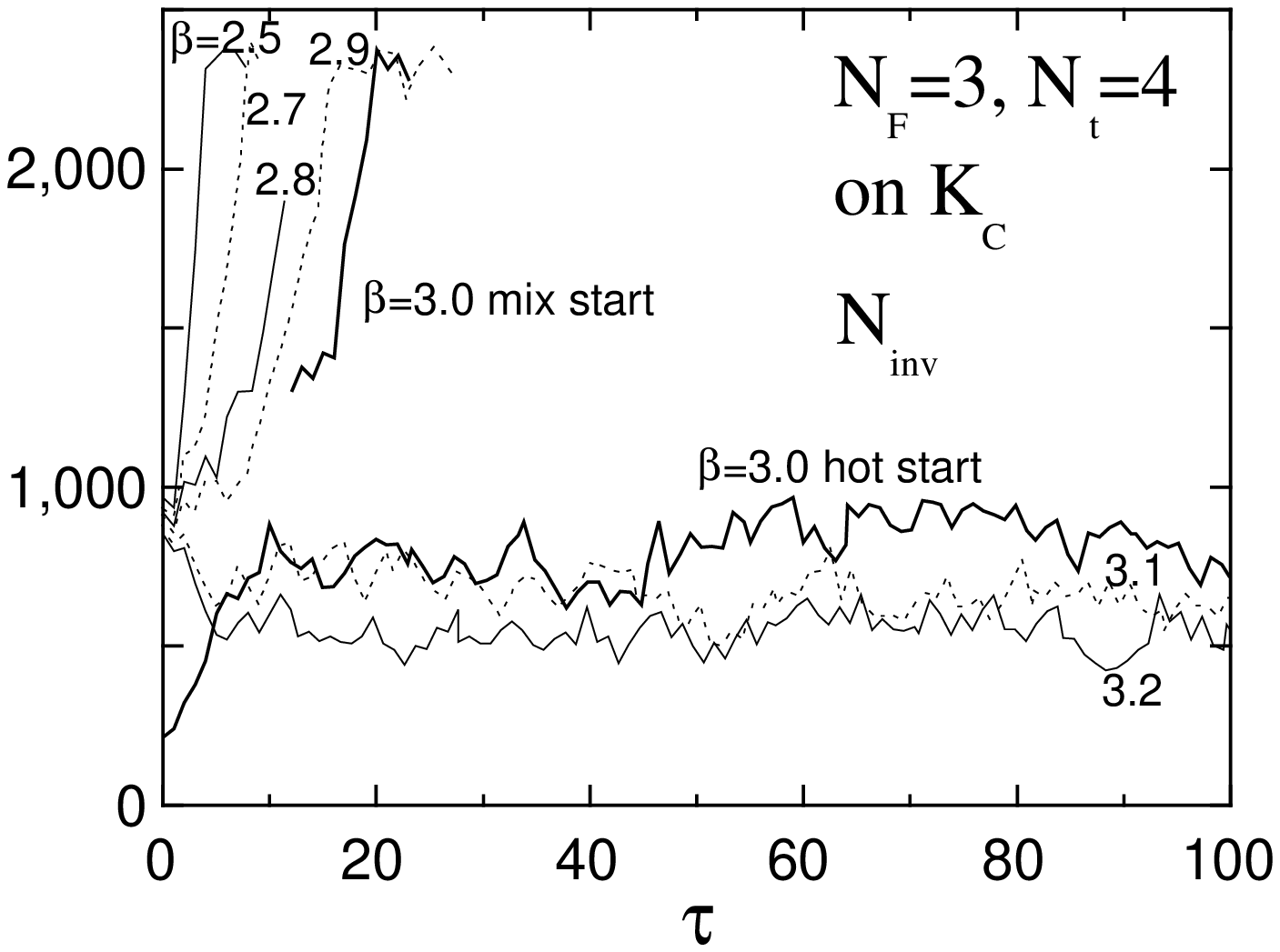}}
\caption{{
Time history of $N_{\rm inv}$ for 
$N_F=3$ on $K_c$. 
}
\label{figH3Ninv}}
\end{figure}

Fig.~\ref{figH3Ninv} shows $N_{\rm inv}$ as a function of the 
molecular-dynamics time $\tau$
for several values of $\beta$'s.
When $\beta \ge 3.1$, $N_{\rm inv}$ is of order of several 
hundreds, while when $\beta \le 2.9$, $N_{\rm inv}$
shows a rapid increase with $\tau$.
At $\beta = 3.0$ we see a clear two-state signal
depending on the initial condition:
For a hot start, $N_{\rm inv}$
is quite stable around $\sim 800$ and $m_\pi^2$ is large 
($\sim 1.0$).
On the other hand, for a mix start, $N_{\rm inv}$ 
shows a rapid increase with $\tau$ and exceeds
2,000 in $\tau \sim 20$, and in accord with this, $m_\pi^2$ 
decreases with $\tau$. 

\begin{figure}
\centerline{ \epsfxsize=13cm \epsfbox{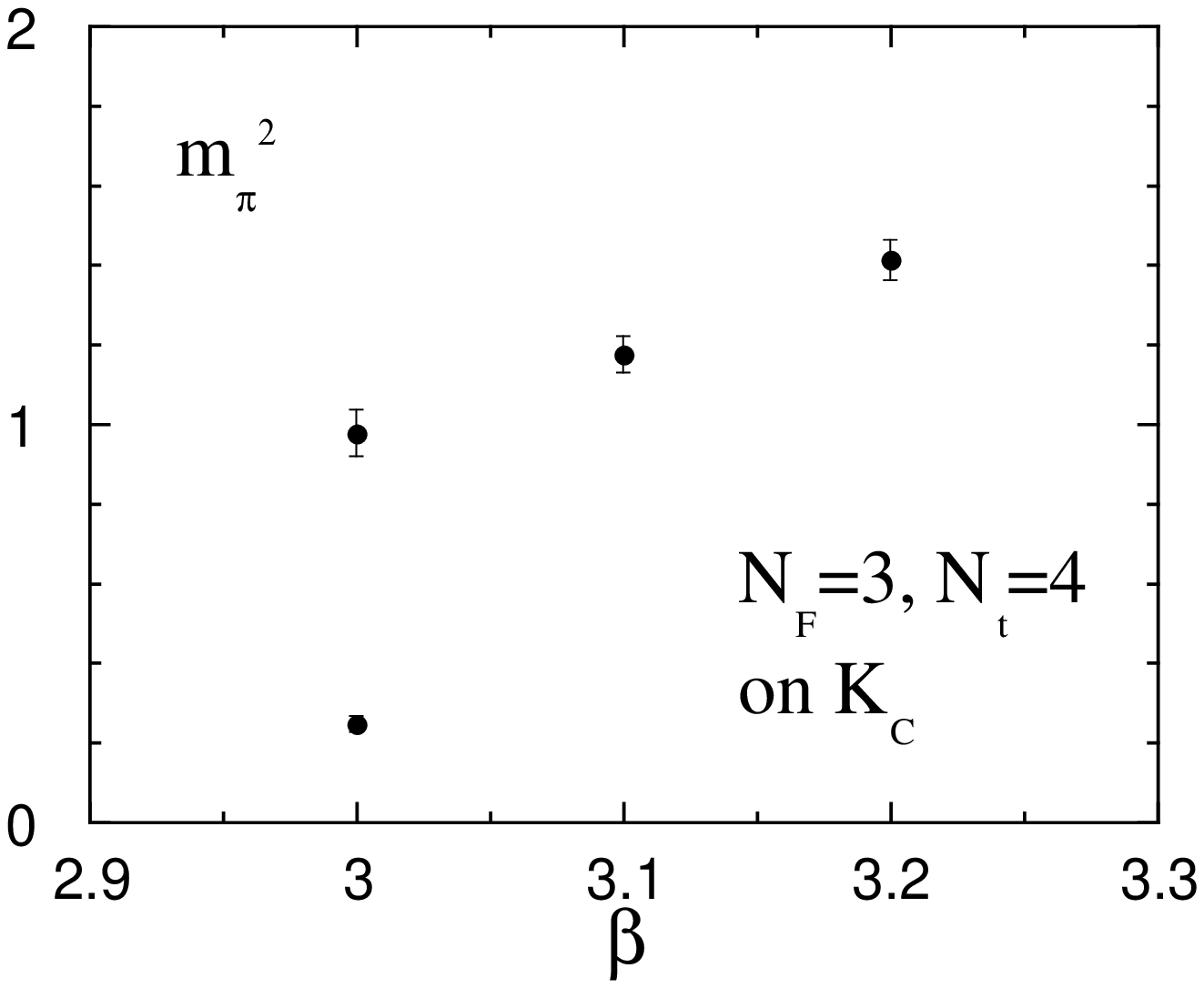}}
\caption{{
$m_{\pi}^2$ for $N_F=3$ on $K_c$. 
}
\label{figF3Mpi}}
\end{figure}

The value of $m_\pi^2$ is plotted in Fig.~\ref{figF3Mpi}.
At $\beta =3.0$ we have two values for $m_\pi^2$ depending 
on the initial configuration. 
The larger one obtained for the hot start
is of order 1.0, which is a smooth extrapolation of the
values at $\beta \sim 3.1$ - 3.2.
The smaller one is an upper bound for $m_\pi^2$ for the 
mix start.

We note that the result of
$\beta_{ct} \sim 3.0$ is consistent with an extrapolation of 
$K_t$ points listed in Table \ref{tab:3} 
as is shown in Fig.~\ref{figF2F3KcKt}.
Thus we identify the crossing point at 
$\beta_{ct} \sim 3.0(1)$.
With the clear two-state signal we 
conclude that the chiral transition is of first order for 
$N_F=3$.

\subsection{$N_F$ = 6}
\begin{figure}
\centerline{ \epsfxsize=13cm \epsfbox{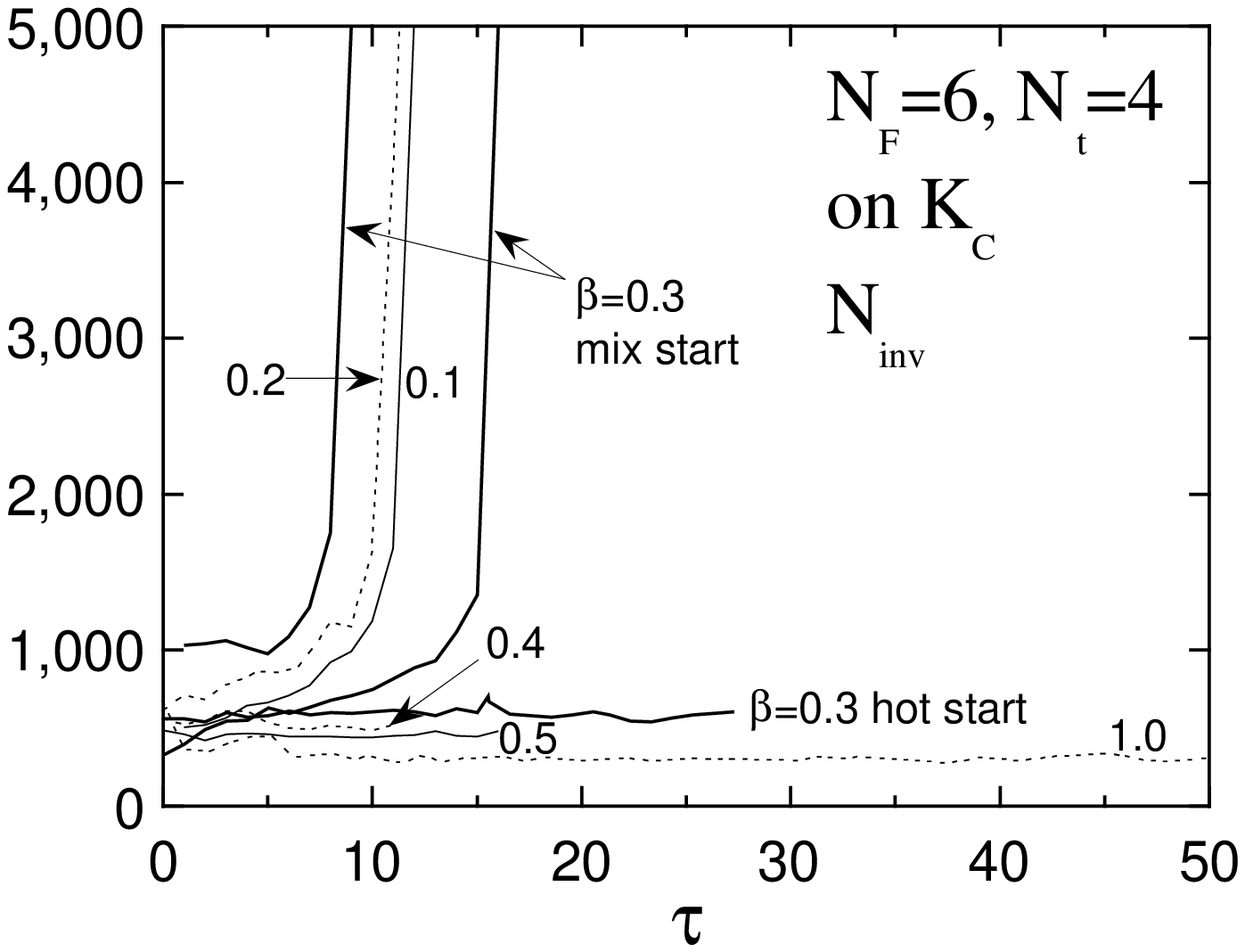}}
\caption{{
Time history of $N_{\rm inv}$ for 
$N_F=6$ on $K_c$. 
}
\label{figH6Ninv}}
\end{figure}

\begin{figure}
\centerline{ \epsfxsize=13cm \epsfbox{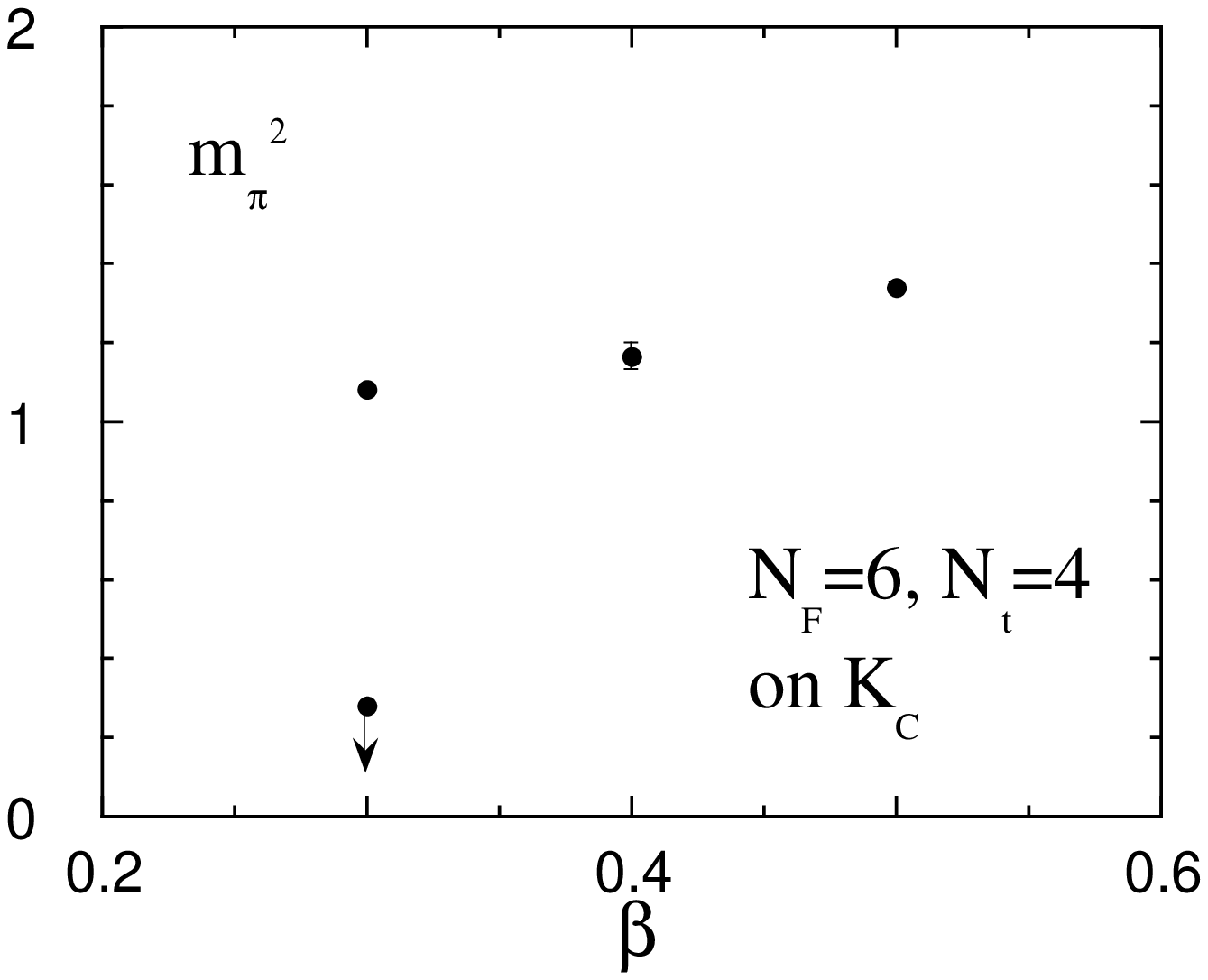}}
\caption{{
$m_{\pi}^2$ for $N_F=6$ on $K_c$. 
}
\label{figF6Mpi}}
\end{figure}

Finally let us discuss the results for $N_F=6$.
Our previous study\cite{strong} implies 
that this number of flavor is critical for
the existence of the crossing point
and therefore it is important to establish the existence 
of the crossing in this case.
Overall features of the transition for $N_F=6$
are very similar to those for $N_F=3$ except for the
position of $\beta_{ct}$, which moves to a smaller 
$\beta$ as expected.
Fig.~\ref{figH6Ninv} shows that $N_{\rm inv}$ ``on-$K_c$''
stays at several hundreds for $\beta \ge 0.4$
and for a hot start at $\beta=0.3$.
On the other hand,
$N_{\rm inv}$ grows rapidly with $\tau$ and
exceeds 5,000 for $\beta \le 0.2$ and for a mix start 
at $\beta=0.3$.
In accord with this, we have two values of $m_\pi^2$ 
at $\beta=0.3$ (Fig.~\ref{figF6Mpi}).
Therefore we identify the crossing point at 
$\beta_{ct} \sim 0.3(1)$
and conclude that the chiral transition is of first order
for $N_F=6$.
This $\beta_{ct}$ is consistent with a linear extrapolation 
of the $K_t$ line ($K_t=0.245$ -- 0.2475, 0.235 -- 0.237, 
and 0.166 -- 0.168 for $\beta=0.5$, 1.0, and 4.5, 
respectively).

\section{Conclusions}
It should be noted that our main results obtained in this 
work that the chiral transitions for $N_F=3$ and 6 are 
of first order,
while it is continuous for $N_F=2$,
are consistent with the prediction based on 
universality\cite{Wilczek}.
Our results with Wilson fermions are also consistent 
with those with staggered fermions\cite{KSfl}.


The simulations 
on the $N_t=4$ and 6 lattices and the $N_t=18$ lattice
have been performed with HITAC S820/80 at KEK 
and with QCDPAX at the University of Tsukuba, respectively.
We would like to thank members of KEK
for their hospitality and strong support
and the other members of QCDPAX collaboration for their help.
This project is in part supported by the Grant-in-Aid
of Ministry of Education, Science and Culture
(No.62060001 and No.02402003).



\clearpage

\begin{table}
\begin{center}
\begin{tabular}{c|cc|cc}
\hline
 & \multicolumn{2}{c|}{$N_t=4$} 
 & \multicolumn{2}{c}{$N_t=8$}\\
$N_F$ & $K_c(m_{\pi}^2)$ & $K_c(m_q)$ 
& $K_c(m_{\pi}^2)$ & $K_c(m_q)$\\
\hline
2 & 0.214(1) & 0.210(1) & 0.212(1) & 0.209(1) \\
3 & 0.209(1) & 0.204(1) &          &          \\
6 & 0.205(2) & 0.200(1) &          &          \\
\hline
\end{tabular}
\caption{
Chiral limit $K_c$ at $\beta=4.5$ 
determined by vanishing $m_{\pi}^2$ and $m_q$. 
Values of $m_{\pi}^2$ and $m_q$ in the confinig phase are 
linearly extrapolated in $1/K$ 
using data from $K=0.16$ -- 0.18 for $N_F=2$ and 3 and
$K=0.15$ -- 0.165 for $N_F=6$ 
(because $K_t = 0.167(1)$ for $N_F=6$ at $N_t=4$). 
The spatial lattice size is $8^2\times 10$.
\protect\label{tab:1}}
\end{center}
\end{table}


\begin{table}
\begin{center}
\begin{tabular}{c|cccc}
\hline
$\beta$          & 3.0 & 3.5 & 4.0 & 4.3\\
\hline
$K_c(m_{\pi}^2)$ & 0.235(1) & 0.230(1) & 0.223(1) & 0.218(1)\\
$K_c(m_q)$       & 0.230(1) & 0.226(1) & 0.218(4) & 0.214(1)\\
\hline
\end{tabular}
\caption{
$K_c$ for $N_F=2$ determined on an $8^2\times10\times4$ 
lattice. 
The results for $\beta=4.5$ are given in Table 
\protect\ref{tab:1}.
\protect\label{tab:2}}
\end{center}
\end{table}


\begin{table}
\begin{center}
\begin{tabular}{cc|cc}
\hline
\multicolumn{2}{c|}{$N_F=2$} & \multicolumn{2}{c}{$N_F=3$}\\
$\beta$ & $K_t$ & $\beta$ & $K_t$\\
\hline
4.3  & 0.207 -- 0.210 & 3.0  & $>$ 0.230      \\
4.5  & 0.200 -- 0.202 & 4.0  & 0.200 -- 0.205 \\
5.0  & 0.170 -- 0.180 & 4.5  & 0.186 -- 0.189 \\
5.25 & 0.160 -- 0.165 & 4.5* & 0.186 -- 0.189 \\
     &                & 4.7* & 0.179 -- 0.180 \\
     &                & 5.0  & 0.166 -- 0.1665 \\
     &                & 5.5  & 0.1275 -- 0.130 \\
\hline
\end{tabular}
\caption{
Finite temperature transition/crossover $K_t$ for $N_F=2$ and 3
obtained on an $8^2\times10\times4$ lattice
(data with * obtained on a $12^3\times4$ lattice). 
\protect\label{tab:3}}
\end{center}
\end{table}

\clearpage

\begin{thebibliography}{9}

\bibitem{Bo}
M.\ Bochicchio {\it et al.},
Nucl.\ Phys.\ B262 (1985) 331.

\bibitem{ItohNP}
S.\ Itoh, Y.\ Iwasaki, Y.\ Oyanagi and T.\ Yoshi\'e,
Nucl.\ Phys.\ B274(1986) 33.
\bibitem{mm}
 L.\ Maiani and G.\ Martinelli,
 Phys.\ Lett.\ 178B (1986) 265.
\bibitem{ChQ}
Y.\ Iwasaki, T.\ Tsuboi and T.\ Yoshi\'e,
Phys.\ Lett.\ B220 (1989) 602.
\bibitem{ChNf2}
Y.~Iwasaki, K.~Kanaya, S.~Sakai and T.~Yoshi\'e,
Phys.\ Rev.\ Lett.\ 67 (1991) 1494.
\bibitem{FOU}
      M.\ Fukugita, S.\ Ohta and A.\ Ukawa,
      Phys.\ Rev.\ Lett.\ 57 (1986) 1974.
\bibitem{strong}
Y.\ Iwasaki, K.\ Kanaya, S.\ Sakai and T.\ Yoshi\'e,
Phys.\ Rev.\ Lett.\ 69 (1992) 21.
\bibitem{preliminary}
Y.\ Iwasaki, K.\ Kanaya, S.\ Sakai and T.\ Yoshi\'e,
Nucl.\ Phys.\ B (Proc.\ Suppl.) 30 (1993) 327;
{\it ibid.} 34 (1994) 314.
\bibitem{zero_mode}
S.\ Itoh, Y.\ Iwasaki and T.\ Yoshi\'e,
Phys.\ Rev.\ D36(1986) 527.
\bibitem{various}
R.\ Gupta {\it et al.}, Phys.\  Rev.\ D40 (1989) 2072;
A.\ Ukawa, Nucl.\ Phys.\  B (Proc.\ Suppl.) 9 (1989) 463;
K.\ Bitar {\it et al.}, Phys.\ Lett.\ B234 (1990) 333;
Phys.\ Rev.\ D43 (1991) 2396;
C.\ Bernard {\it et al.}, {\it ibid.} D46 (1992) 4741;
{\it ibid.} D49 (1994) 3574;
Nucl.\ Phys.\ B (Proc.\ Suppl.) 34 (1994) 324;
T. Blum {\it et al.},
Phys.\ Rev.\ D50 (1994) 3377.
\bibitem{kc}
See references in Y.\ Iwasaki, preprint of Tsukuba 
UTHEP-293(1994).
\bibitem{Wilczek} R.\ Pisarski and F.\ Wilczek,
Phys.\ Rev.\ D29 (1984) 338;
F.\ Wilczek, Int.\ J.\ Mod.\ Phys.\ A7 (1992) 3911; 
K.\ Rajagopal and F.\ Wilczek, Nucl.\ Phys.\ B399 (1993) 395.
\bibitem{KSfl}
For recent review, C.\ DeTar, preprint of Utah, UU-HEP94/4; 
F.\ Karsch, 
Nucl.\ Phys.\ B (Proc.\ Suppl.) 34 (1994) 63.

\end{thebibliography}
\end{document}